\documentclass[useAMS,usegraphicx]{mn2e}
\usepackage{rotate}
\usepackage{rotating}
\usepackage{times}
\newif\ifAMStwofonts
\AMStwofontstrue

\def\gs{\mathrel{\hbox{\rlap{\hbox{\lower4pt\hbox{$\sim$}}}\hbox{$>$}}}}
\def\ls{\mathrel{\hbox{\rlap{\hbox{\lower4pt\hbox{$\sim$}}}\hbox{$<$}}}}

\def\suzaku{{\it Suzaku}}

\def\swift{{\it Swift}}
\def\xmm{{\it XMM-Newton}}

\def\et{{et al.\ }}

\def\mrk335{{Mrk~335}}
\def\rg{{\thinspace r_{\rm g}}}

\def\fvar{{F_{\rm var}}}
\def\chidof{{\chi^2_\nu/{\rm dof}}}
\def\redchi{{\chi^2_\nu}}

\def\feka{{Fe~K$\alpha$}}
\def\fela{{Fe~L$\alpha$}}

\def\fekb{{Fe~K$\beta$}}

\def\fexxvi{{Fe~\textsc{xxvi}}}
\def\nh{{N_{\rm H}}}

\def\deg{^{\circ}}

\def\cm{{\rm\thinspace cm}}
\def\erg{{\rm\thinspace erg}}
\def\eV{{\rm\thinspace eV}}

\def\ga{{\rm\thinspace gauss}}

\def\keV{{\rm\thinspace keV}}
\def\km{{\rm\thinspace km}}

\def\s{{\rm\thinspace s}}
\def\ks{{\rm\thinspace ks}}
\def\ps{{\rm\thinspace s^{-1}}}

\def\cts{{\rm\thinspace count}}

\def\cps{\hbox{$\cts\s^{-1}\,$}}
\def\cmps{\hbox{$\cm\s^{-1}\,$}}

\def\ergpscmps{\hbox{$\erg\cm^{-2}\s^{-1}\,$}}

\def\kmps{\hbox{$\km\ps\,$}}

\def\pscm{\hbox{$\cm^{-2}\,$}}

\title[The blurred reflection interpretation for \mrk335]
      {
A blurred reflection interpretation for the intermediate flux state in \mrk335\        }

\author[L. C. Gallo et al.]
       {L. C. Gallo,$^1$ 
       A. C. Fabian,$^2$
	D. Grupe,$^3$
	K. Bonson,$^1$
	S. Komossa,$^4$
	A. L. Longinotti,$^5$
	\newauthor	
	G. Miniutti,$^6$  
	D. J. Walton,$^{2,7}$
 A. Zoghbi$^8$
and S. Mathur$^9$ 
        \\ 
$^{1}$ Department of Astronomy and Physics, Saint Mary's University, 923 Robie Street, Halifax, NS, B3H 3C3, Canada \\
$^{2}$ Institute of Astronomy, University of Cambridge, Madingley Road, Cambridge CB3 0HA\\
$^{3}$ Department of Astronomy and Astrophysics, Pennsylvania State
University, 525 Davey Lab, University Park, PA 16802, USA \\
$^{4}$  Max-Planck-Institut f\"ur Radioastronomie, Auf dem H\"ugel 69, 53121 Bonn, Germany \\
$^{5}$ XMM-Newton Science Operations Centre, ESA, Villafranca del
Castillo, Apartado 78, E-28691 Villanueva de la Ca\~nada, Spain \\
$^{6}$ Centro de Astrobiologia (CSIC-INTA), Dep. de Astrofisica; LAEFF, PO Box 78, E-28691, Villanueva de la Ca\~nada, Madrid, Spain \\
$^{7}$ California Institute of Technology, 1200 East California Boulevard, Pasadena, CA 91125, USA  \\
$^{8}$ Department of Astronomy, University of Maryland, College Park, MD 20742, USA \\
$^{9}$ Astronomy Department, The Ohio State University, Columbus, OH 43210, USA \\
}
\date{Accepted. Received. }
\pagerange{\pageref{firstpage}--\pageref{lastpage}}
\pubyear{2010}
\begin{document}
\maketitle
\label{firstpage}

\begin{abstract}
As part of a long term monitoring campaign of \mrk335, 
deep  \xmm\  observations catch the narrow-line Seyfert 1 galaxy (NLS1) in a complex, intermediate flux interval as the active galaxy is transiting from low- to high-flux.  Other works on these same data examined the general behaviour of the NLS1 (Grupe \et) and the conditions of its warm absorber 
(Longinotti \et).
The analysis presented here demonstrates the  X-ray continuum and timing properties  can be described in a self-consistent manner adopting a blurred reflection model with no need to invoke partial covering.  The rapid spectral variability appears to be driven by changes in the shape of the primary emitter that is illuminating the inner accretion disc around a rapidly spinning black hole ($a>0.7$).   While light bending is certainly prominent, the rather constant emissivity profile and break radius obtained in our spectral fitting suggest that the blurring parameters are not changing as would be expected if the primary source is varying its distance from the disc. Instead changes could be intrinsic to the power law component.  One possibility is that material in an unresolved jet above the disc falls to combine with material at the base of the jet producing the changes in the primary emitter (spectral slope and flux) without changing its distance from the disc.

\end{abstract}



\section{Introduction}
\label{sect:intro}

The narrow-line Seyfert 1 galaxy (NLS1), \mrk335 ($z=0.025785$), has always been the subject of intense scrutiny.  As one of the brighter active galactic
nuclei (AGNs) in the X-ray sky it has been observed by most major missions so its long term behaviour is relatively well established (Grupe \et 2008).  
Its X-ray continuum has normally been described with a power law and soft-excess (e.g. Pounds \et 1987).  More sensitive observations revealed a narrow emission line around $6.4\keV$ along with a broader emission feature in the \feka\ region.  The combination of the soft excess and broad emission feature is indicative of reflection from the inner region of an ionised disc (e.g. Ross \& Fabian 2005) and such models have been successful in fitting the spectra of \mrk335\ (e.g. Ballantyne, Iwasawa \& Fabian 2001; Crummy \et 2006; O'Neill \et 2007; Longinotti \et 2007; Larsson \et 2008).

A 2007 \swift\ snap-shot of \mrk335\ found the NLS1 in an unprecedented low state where the X-ray flux had diminished to about 1/30 of the previously lowest observed flux state (Grupe \et 2007).  The \swift\ observation showed considerable curvature in the X-ray spectrum, and simultaneous optical data found that the AGN was in an X-ray weak state.  Both factors are indicative of either absorption or reflection dominating the spectrum (Gallo 2006).  The \swift\ observation prompted a $20\ks$ \xmm\ Target of Opportunity observation that revealed low-energy emission lines from a photoionised gas likely originating far from the inner accretion disc and close to the broad line region (Grupe \et 2008; Longinotti \et 2008).  During this low state the continuum could be described equally well as arising from either partial covering or blurred reflection (Grupe \et 2008).  

Since 2007 \mrk335\ has undergone continuous monitoring with \swift.  The AGN has exhibited significant UV and X-ray variability while remaining in a moderately low X-ray flux state.  A deep $200\ks$ pointed \xmm\ observation in 2009 attempted to catch the AGN in a minimum in order to study the properties of the photoionised gas observed in the 2007 low state; however, the AGN was instead caught in a transition as it brightened to an intermediate flux level (Grupe \et 2012, hereafter G12).  Interestingly, while the emitting gas is less evident in this intermediate state due to the higher continuum flux compared to the 2007 low state, a warm absorber is now clearly detected (G12; Longinotti \et 2012, hereafter L12).  Based on their analysis of the RGS data, L12 describe the warm absorber as a combination of three different zones with distinct temperatures and ionisation parameters.   However, the RGS analysis shows that the ionized gas does not produce the variability observed during this intermediate flux state.  The curvature in the X-ray spectrum again could be described by partial covering prompting the suggestion that clouds of absorbing material have been entering the line of sight since 2007 and variations in covering fraction are responsible for the recent low state and variability exhibited by \mrk335 (G12).  Grupe \et (2008) point to Tanaka \et (2005) for a partial covering explanation of \mrk335\ in the high-flux state when the covering fraction would have been low.

If the recent diminished flux is due to increased photoelectric absorption from additional partial covering, 
then it should 
be accompanied by increased emission in the narrow \feka\ line (e.g. Reynolds \et 2009).
G12 find that within uncertainties the \feka\ flux in \mrk335\ is identical in the intermediate- and low-flux states, 
both of which are comparable (perhaps even lower) than the \feka\ flux reported by Grupe \et (2008) in the high-flux state.
In addition, one may expect to find markers of absorption in the UV.  There is some indication of absorption in a recent HST-COS
spectrum of \mrk335\ (L12).  However if and how this is linked to any absorption in the X-ray is difficult to establish since the UV and X-ray observations were not simultaneous.

In this work we test the blurred reflection interpretation for \mrk335\ when the AGN is in an intermediate low-flux
state and is significantly modified by a warm absorber, to determine if partial covering is necessary to describe the 
behaviour of the source in recent years.
The observations and data reduction
are described in the following section.  Spectral modelling is explored in Section~3 and the temporal analysis is described in Section~4.   We discuss and summarize our results in Section~5 and 6, respectively.

\section{Observations and data reduction}
\label{sect:data}
\mrk335\ has been observed with \xmm\ (Jansen \et 2001) on five occasions between 2000 and 2009.
In this work we report on the long 200~ks observation spread over two consecutive orbits (revolutions 1741 and 1742) in 2009.
During the observations the EPIC pn detector (Str\"uder \et 2001) was operated in large-window mode and with the thin filter in place. The MOS (MOS1 and MOS2; Turner \et 2001) cameras were operated in full-frame mode and with the medium filter.  The Reflection Grating Spectrometers (RGS1 and RGS2; den Herder \et 2001) also collected data during these observations, as did the Optical Monitor (OM; Mason \et 2001).  The OM data have been presented in G12 and will not be reanalyzed in this work.  These data will be discussed as needed.  The RGS analysis is the subject of an in-depth study of the warm absorber (L12).  We will incorporate the results of L12 in this work, but we will not reanalysis the RGS data here.

The \xmm\ Observation Data Files (ODFs) from each observations
were processed to produce calibrated event lists using the \xmm\ 
Science Analysis System ({\tt SAS v11.0.0}).
Unwanted hot, dead, or flickering pixels were removed as were events due to
electronic noise.  Event energies were corrected for charge-transfer
inefficiencies.  
Light curves were extracted from these
event lists to search for periods of high background flaring and such  
periods have been neglected.
Pile-up was negligible during the observations.
The background photons were extracted from an off-source region on the same CCD.
Single and double events were selected for the pn detector, and
single-quadruple events were selected for the MOS.
EPIC response matrices were generated using the {\tt SAS}
tasks {\tt ARFGEN} and {\tt RMFGEN}.  
The MOS and pn data at each epoch were compared for consistency and determined to
be in agreement within known uncertainties (Guainazzi \et 2010) as was the case in G12.  
For simplicity we will only present the pn data at each epoch.  The total amount of good exposure for the pn during revolution 1741 and 1742 is  96.4, 
and $70.0\ks$, respectively.

The \xmm\ X-ray spectra were grouped such that each bin contained at least 30 counts.
Spectral fitting was performed using {\tt XSPEC v12.7.1}
(Arnaud 1996).
All parameters are reported in the rest frame of the source unless specified
otherwise.  Figures remain in the observed frame and are binned for display purposes only.
Unless stated otherwise, black, open symbols identify data from revolution 1741; and filled, red squares represent data from orbit 1742.
The quoted errors on the model parameters correspond to a 90\% confidence
level for one interesting parameter (i.e. a $\Delta\chi^2$ = 2.7 criterion).
A value for the Galactic column density toward \mrk335\ of
$3.56 \times 10^{20}\pscm$ (Kalberla \et 2005) is adopted in all of the
spectral fits and
abundances are from Anders \& Grevesse (1989).
Luminosities are calculated using a
Hubble constant of $H_0$=$\rm 70\ km\ s^{-1}\ Mpc^{-1}$ and
a standard flat cosmology with $\Omega_{M}$ = 0.3 and $\Omega_\Lambda$ = 0.7.


\section{Spectral Modeling}
\subsection{The high-energy spectral region}
\label{sect:high}

The $2.5-10\keV$ band in \mrk335\ typically displays curvature both in the high- and low-flux
states that can be attributed to blurred reflection (e.g. O'Neill \et 2007; Larsson \et 2008; Grupe \et 2008) or partial covering (e.g. Grupe \et 2008; G12).  Narrow features have also been detected.  A narrow $\sim6.4\keV$ emission line arising from \feka\ is
regularly observed, while narrow emission features from a highly ionized plasma (\fexxvi\ Ly-$\alpha$, O'Neill \et 2007) or absorption features attributed to relativistic inflows (Longinotti \et 2007) are transient.
\begin{figure}
\rotatebox{270}
{\scalebox{0.32}{\includegraphics{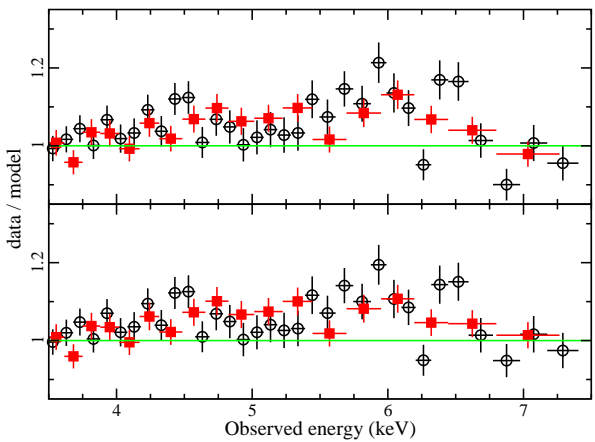}}}
\caption{
The $2.5-10\keV$ spectrum is fitted at each epoch with slightly different models for the distant reflector. 
Top panel: Residuals (data/model) remaining in the $3.5-7.5\keV$ after fitting the spectra with a power law plus {\tt pexmon}
model.  
Lower panel:  Same as above, but using {\tt reflionx} to model the narrow \feka\ emission line. In both figures the black open circles correspond to the data from revolution 1741 and the filled, red squares are data from revolution 1742.
This colour/symbol convention is used throughout the paper unless stated otherwise.
}
\label{fig:resid}
\end{figure}

The $2.5-10\keV$ spectrum from each epoch was fitted with a power law and a second component to mimic distant reflection.  For the distant reflector we consider two models, {\tt pexmon} (Nandra \et 2007) and {\tt reflionx} (Ross \& Fabian 2005).  Both models self-consistently include \feka, the Compton shoulder, and Compton hump.  
In addition, {\tt pexmon} includes \fekb\ and Ni~K$\alpha$ emission.    However, {\tt pexmon} is for reflection off neutral material and does not consider the features that arise below the iron-K band.  For  {\tt reflionx}  the material is also considered nearly neutral as the 
ionisation parameter ($\xi = 4\pi F_x /n$, where $F_x$ is the incident flux and $n$ is the density of the gas) 
was set to $1 \erg\cmps$ (the lowest allowed by the model).   Iron abundance is also fixed to the solar value.  While there is evidence that the inner accretion disc can show super-solar iron abundances (e.g. Fabian \et 2009; Reynolds \et 2012) it is not obvious that this characteristic extends to the distant reflector.
However, we note that Walton \et (2012a) was able to model the \suzaku\ spectrum of \mrk335\ when they allowed the 
distant reflector (torus) and the inner accretion disc to share common abundances.

Statistically the models fit reasonably well ($\chidof=1.03/1507$ and $1.01/1508$ for the {\tt pexmon} and {\tt reflionx}, respectively), but both models show excess in the residuals between $3.5-7.5\keV$ (Fig.~$\ref{fig:resid}$).  G12 interpreted this excess as evidence for partial covering, we examine this in terms of blurred, ionised reflection from the inner accretion disc.  Adding a single Laor profile to both models improved the fits ($\redchi\ls1.0$) and residuals significantly in both cases.  The equivalent width of the line is $EW\sim400\eV$ in both models.  The fitted line parameters are the energy ($E\approx6.4\keV$), disc emissivity index ($q\approx4-8$), inner disc radius ($R_{in}\approx 1.3-5\rg$, [$\rg= GM/c^{2}$]), and disc inclination ($i\approx 30-65\deg$).  We do not examine these parameters in detail at this point as they will serve as initial values for more realistic models in the remaining analysis.
 

\subsection{The mean $0.3-12\keV$ spectrum}
\label{sect:bestfit}

\begin{figure}
\rotatebox{270}
{\scalebox{0.32}{\includegraphics{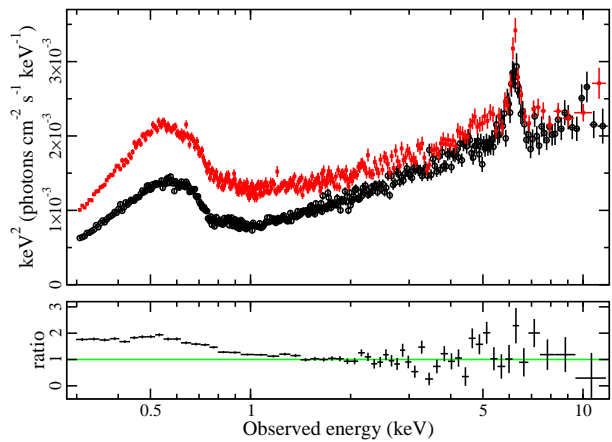}}}
\caption{
Top panel: The average spectrum in revolution 1741 (black circles) and revolution 1742 (red squares).  
Lower panel:  The ratio of the difference spectrum (rev. 1742 -- rev. 1741) fitted with an absorbed power law ($\Gamma\approx2.5$) between $2-10\keV$ and extrapolated to lower energies.  As seen in the top panel the variability is dominate below $1.5\keV$.
}
\label{fig:diff}
\end{figure}

As seen in Fig.~\ref{fig:diff} the average spectrum undergoes significant changes between revolution 1741 and 1742.  The difference between the two
spectra (i.e. rev. 1742 -- rev. 1741) fitted with an absorbed power law ($\Gamma\approx2.5$) between $2-10\keV$ and extrapolated to lower energies is shown in the lower panel of  Fig.~\ref{fig:diff}.  The changes are substantial below $\sim1.5\keV$, but there is also indication for variability above $2\keV$.  For this reason we will treat the data from each epoch separately when examining the spectra.

There is a significant warm absorber present at both epochs, which is clearly seen in the RGS spectra, and is the subject of other work (L12).  This is the first time that a warm absorber has been unambiguously observed in \mrk335.  L12 identified three distinct ionized components, all outflowing at high velocity ($\sim5000\kmps$).  However, since L12 determined that the warm absorber itself is not variable during these observations, 
it cannot be the warm absorber that is responsible for the changes seen in Fig.~\ref{fig:diff}.
To remain consistent with the RGS analysis, we constructed {\tt XSTAR} (Kallman \& Bautista 2001) grids that are comparable with the three warm absorbers found in L12.  These components are included in our spectral fitting and are linked between the epochs.

The neutral reflector originating from distant matter is modelled with {\tt reflionx} rather than {\tt pexmon} since it offers a more complete treatment of the reflection at low energies.  This component should not be confused with the reflection from the inner disc as it is not subject to any blurring effects.  The ionisation parameter of this distant reflector is fixed at $\xi=1.0\erg\cmps$ (i.e. the lowest value permitted by the model) and the iron
abundance is fixed to solar.  The photon index of the ionising power law is fixed to the canonical value of $\Gamma=1.9$ (Nandra \& Pounds 1994).  It is not clear if this assumption is accurate as the photon index is clearly variable in \mrk335\ (e.g. $1.5\ls\Gamma\ls2.3$ have been reported).   It is also not obvious that the torus would be illuminated by the same power law that is seen by the observer since the line-of-sight through the corona, and hence the optical depth is likely different from each perspective.  The normalisation of this component is linked at both epochs.

In the 2007 low flux state observation, Grupe \et (2008) and Longinotti \et (2008) identified narrow emission features attributed 
to a distant photoionised plasma.  The CCD spectra were fitted with a {\tt MEKAL} model to account for this emission (Grupe \et 2008), but the fitting was rather ad hoc and did not account for some of the strongest features in the
spectrum (e.g. a $0.89\keV$ emission feature and the narrow \feka\ emission line).  Such emission is not prominent in the 
2009 RGS data (L12) and this could simply be a matter of the higher continuum flux overwhelming the hot, diffuse plasma.
We do not specifically consider the reasonable assumption that a constant {\tt MEKAL} component (e.g. fixed to the 2007 values) 
is present in these 2009 data, but we note that the distant reflector modelled by the unblurred {\tt reflionx} component can 
similarly describe this emission.  The $0.5-10\keV$ flux of the {\tt MEKAL} component in the low-flux spectrum was $\sim2.2\times10^{-13}\ergpscmps$, comparable to the flux of the distant reflector used in these models (see Table~\ref{tab:fits}). 

After fitting the three-phase warm absorber to the RGS data, L12 attributed the remaining residuals between about $0.49-0.65\keV$
to intrinsically narrow emission corresponding to {\sc O viii} and {\sc N vii} Ly-$\alpha$, as well as the {\sc O vii} triplet.
Adding Gaussian profiles with fixed energies and widths, but free normalisations, to the CCD data did not result in any detection
of these features.  For each line, the measured equivalent width was $EW<1\eV$.
Including the Gaussian features with all parameters fixed to the RGS measured values generated an entirely comparable fit to the
one presented in Table~\ref{tab:fits}.  

The broadband continuum is described with a power law plus reflection from the accretion disc.  
It is then modified by a three-zone warm absorber and a distant neutral absorber. 
The reflection component is blurred for dynamical effects in the disc close to the black hole using {\tt kdblur2} with a broken power law emissivity profile.   The objective was to obtain a self-consistent model that described both observations.  Despite the significant variability the observations were separated by only $\sim 42\ks$, consequently we were selective in what we allowed to vary between the observations.  In addition to the distant reflector, the inclination ($i$), iron abundance ($A_{Fe}$), inner ($R_{in}$) and outer ($R_{out}$) disc radius were linked between the observations.  Initial fits were conducted with the inner emissivity profile ($q_{in}$) and the break radius ($R_{b}$) free to vary at each epoch.  The outer emissivity profile ($q_{out}$) was fixed to 3 as would be expected in the case for illumination of the disc from a distant source.  Finding no significant differences in the values of $R_{b}$ and $q_{in}$ during revolution 1741 and 1742, the parameters were also linked in subsequent fits.
The best-fitting model reported here allow only the photon index and normalisation of the power law component; and the ionisation and normalisation of the reflector to vary at each epoch.   

\begin{table*}
\caption{The best-fit model parameters for the multi-epoch spectral modelling of \mrk335.  
The model components and model parameters are listed in Columns 1, and 2, respectively. 
Columns 3 and 4 list the parameter values during revolution 1741 and 1742, respectively.
The inner ($R_{in}$) and outer ($R_{out}$) disc radius are given in units of gravitational radii 
($1\rg = GM/c^2$).  The reflection fraction ($\mathcal{R}$) is approximated as the ratio of the reflected flux over
the power law flux in the $20-50\keV$ band. 
Values that are linked between epochs appear only in column 3.  
The superscript $f$ identifies parameters that are fixed and the superscript $p$ indicates values that are pegged at a limit.
Fluxes are corrected for Galactic absorption and are reported in units of $\ergpscmps$.}
\centering
\scalebox{1.0}{
\begin{tabular}{ccccc}                
\hline
(1) & (2) & (3) & (4)  \\
 Model Component &  Model Parameter  &  1741 & 1742  \\
\hline
 Warm Absorber 1 & $\nh$ ($\times10^{21}$) & $2.92^{+0.89}_{-0.62}$ &   \\
           &log$\xi$ [$\erg\cmps$] & $0.98^{+0.05}_{-0.19}$   \\
            Warm Absorber 2 & $\nh$ ($\times10^{21}$) & $1.24^{+0.86}_{-0.54}$  &   \\
           &log$\xi$ [$\erg\cmps$] & $1.89\pm0.08$   \\
            Warm Absorber 3 & $\nh$ ($\times10^{21}$) & $11.8^{+2.9}_{-5.5}$  &   \\
           &log$\xi$ [$\erg\cmps$] & $2.81\pm0.05$   \\
\hline
 Power law & $\Gamma$ & $1.83\pm0.01$ & $1.98\pm0.02$  \\
           &log $F_{0.5-10 keV}$ & $-11.27\pm0.02$ & $-11.24\pm0.04$  \\
\hline
  Blurring  & $q_{in}$    & $7.03^{+1.24}_{-0.85}$   &  \\
            & $R_{in}$ ($\rg$)   & $1.38^{+0.07}_{-0.145p}$        &                                \\
           & $R_{out}$ ($\rg$)   & $400^{f}$          &                                \\
           & $R_b$ ($\rg$)   & $5.78^{+1.94}_{-1.16}$ &  \\
           & $q_{out}$    & $3^{f}$          &   \\
           & $i$ ($\deg$)  & $51^{+10}_{-13}$  &                                      \\
\hline
  Reflection  & $\xi$ ($\erg\cmps$)   & $200^{+2}_{-23}$  & $244^{+27}_{-15}$          \\
             & $A_{Fe}$ (Fe/solar)   & $1.67\pm0.18$           \\
             &log $F_{0.5-10 keV}$ & $-11.32\pm0.02$ & $-11.06\pm0.02$    \\
             & $\mathcal{R}$ & $1.79$ & $2.05$   \\
\hline
  Distant Reflector & $\xi$ ($\erg\cmps$)   & $1.0^{f}$  &          \\
             & $A_{Fe}$ (Fe/solar)   & $1.0^{f}$  &         \\
     & $\Gamma$ & $1.9^{f}$ &  \\  
             &log $F_{0.5-10 keV}$ & $-12.24^{+0.03}_{-0.06}$   \\
             & $\mathcal{R}$ & $1.13$ & $1.59$   \\
\hline
              Fit Quality & $\chidof$ & $1.06/2379$     \\
\hline
\label{tab:fits}
\end{tabular}
}
\end{table*}
\begin{figure}
\rotatebox{270}
{\scalebox{0.32}{\includegraphics{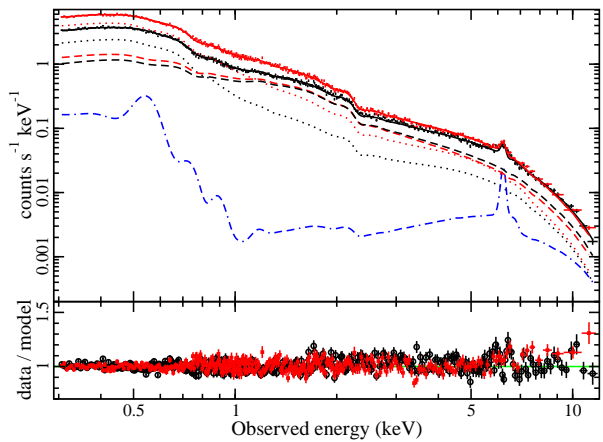}}}
\caption{
Top panel: The average spectrum in revolution 1741 (black circles) and revolution 1742 (red squares) over the $0.3-12\keV$ band fitted with the best-fit model (Table~\ref{tab:fits}).  The dotted lines are the blurred reflector and the dashed curves are the intrinsic power law.  The distant reflector is common to both spectra and is shown in blue (dashed-dotted curve).  
Lower panel:  The data-to-model ratio resulting from the best-fit model.
}
\label{fig:bestfit}
\end{figure}
\begin{figure}
\rotatebox{270}
{\scalebox{0.32}{\includegraphics{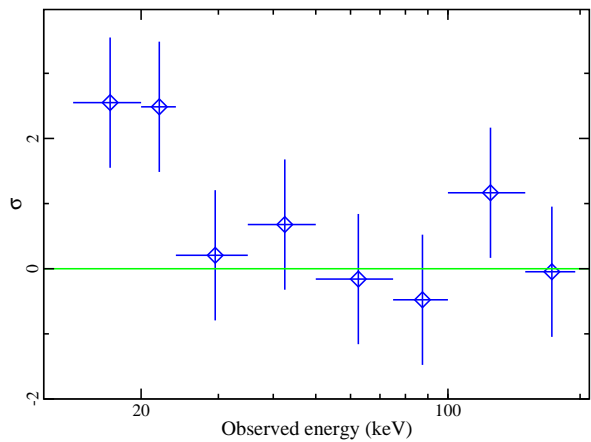}}}
\caption{
The best-fit model described in this section is extrapolated to high energies and compared (i.e. not fitted) to the \swift\ BAT data .
The residuals are shown in terms of sigma.
}
\label{fig:swift}
\end{figure}
The model adequately fits the data ($\chidof=1.06/2379$).  The model parameters and fit are shown in Table~\ref{tab:fits} and Fig.~\ref{fig:bestfit}, respectively.  The reflection model makes specific predictions of the spectral appearance above $10\keV$.  While simultaneous observations above $10\keV$ do not exist, the average \swift\ BAT spectrum is available for comparison (Fig.~\ref{fig:swift}).  The \swift\ data agree well with blurred reflection spectral model.

The differences between the two epoch require changes in the normalisations as well as the shape of the reflection and power law components.  In contrast to  typical observations the reflection component appears more dominant during revolution 1742 when the AGN is in a higher flux state.  

\subsection{Black hole spin}
\begin{figure}
\rotatebox{270}
{\scalebox{0.32}{\includegraphics{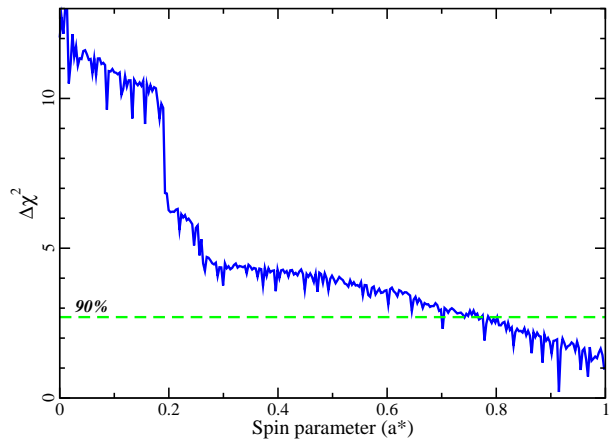}}}
\caption{
The quality of the fit is tested against the black hole spin parameter showing that extremely low spins are discriminated against.  The minimum in $\Delta\chi^2$ is at $a \approx 0.9$.  The green, dashed line marks the 90 per cent confidence level. }
\label{fig:spin}
\end{figure}

In accordance with the small inner radius found in the fit (Table~\ref{tab:fits}) a high black hole spin value is expected in \mrk335.  In order to estimate the black hole spin we conservatively concentrate efforts on the high-energy spectrum above $2.5\keV$.  When fitting the broadband spectrum the spin value is often highly dependent on the interpretation and 
modeling of the smooth soft excess where the statistics are highest, and as we know from L12, 
this region is significantly modified by the three-zone warm absorber.  

We begin with the power law plus distant reflector model described in Section~\ref{sect:high}.  
To this we add a reflector that is blurred using {\tt kerrconv} (Brenneman \& Reynolds 2006; instead of {\tt kdblur2}), 
which includes black hole spin as a model parameter.   As in the broadband spectral analysis, all model parameters are tied except for the power law photon index and normalisation, and the blurred reflector ionisation and normalisation.  A good fit to the $2.5-10\keV$ data is found ($\chidof = 0.96/1307$) with all model parameters in agreement with the broadband results in Table~\ref{tab:fits} (albeit with larger uncertainties).

In order to estimate the uncertainty on the measured spin we examine the variation in the quality of the fit associated with the changes in the spin parameter (Fig.~\ref{fig:spin}).  Relatively high spin values are preferred by the fit ($0.76 < a < 0.998$) at 90 per cent confidence.  Very low spin values ($a < 0.2$) are rejected at the $3\sigma$ level.  The spin parameter measured from these data of \mrk335\ in a warm absorber dominated, intermediate flux state is in agreement with that measured by Walton \et (2012a) using \suzaku\ XIS and PIN  spectra of \mrk335\ in an unabsorbed, high-flux state ($a = 0.8\pm 0.1$), and commensurate with the small inner disc radius found by Crummy \et (2006).

\section{X-ray variability }
\label{sect:xvar}

\begin{figure}
\rotatebox{270}
{\scalebox{0.32}{\includegraphics{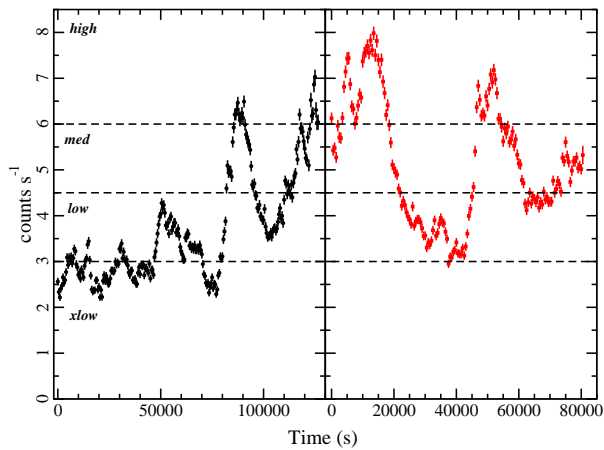}}}
\caption{
The $0.2-12\keV$ pn light curve in $500\s$ bins from revolution 1741 (left) and 1742 (right).  Zero second marks the start of each observation.  The dashed lines delineate the various flux levels examined in Section~\ref{sect:frs}.  The high, medium, and low state are defined with count rates of $> 6$, $4.5-6.0$, and $3-4.5\cps$.  Revolution 1741 also is separated into an extra low state ($<3\cps$).
}
\label{fig:lc}
\end{figure}
\begin{figure}
\rotatebox{270}
{\scalebox{0.32}{\includegraphics{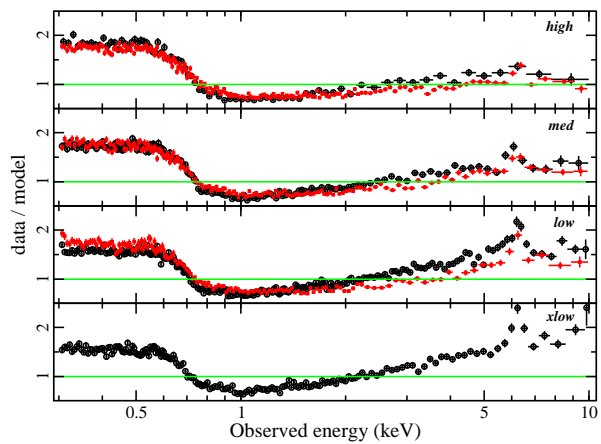}}}
\caption{
Spectra extracted from the flux levels defined in Fig.~\ref{fig:lc} are compared to an absorbed power law with $\Gamma=2$.  Despite the comparable flux there exist differences between the flux-resolved spectra from the two observations. }
\label{fig:states}
\end{figure}
\begin{figure}
\rotatebox{270}
{\scalebox{0.32}{\includegraphics{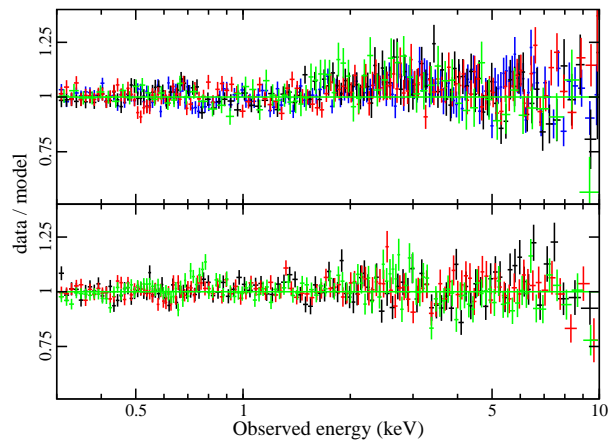}}}
\caption{
The ratio (data / model) from fitting the various flux states defined in Fig.~\ref{fig:lc} with a model in which the reflection and power law component are allowed to vary (see text for details). Data from 1741 and 1742 are shown in the top and bottom panel, respectively.  Black, red, and green data points mark the low-, medium-, and high-flux states.  Blue data points in the top panel identify the extra-low-flux state during 1741.}
\label{fig:fluxed}
\end{figure}

G12 reported significant flux and spectral variability over the intermediate flux level observations examined here (see their Fig.~5).  We attempt to describe the variability in \mrk335\ based on the blurred reflection model presented in Section~\ref{sect:bestfit}.  As noted in Section~\ref{sect:bestfit}, variability is seen in power law normalisation and shape as well as in the ionisation and flux of the reflector.  Here we examine how the model can account for the spectral variability as a function of time and flux level.

\subsection{Flux resolved spectroscopy}
\label{sect:frs}
Spectra are created in four flux bins during revolution 1741 and three flux bins during revolution 1742, as defined in Fig.~\ref{fig:lc}.  The high-, medium-, and low-flux states are defined identically in 1741 and 1742. The extra-low-flux state (i.e. xlow)  is unique to 1741 as the AGN brightness does not dip to this level  during 1742.  In Fig.~\ref{fig:states} we plot the spectra from each flux state against a power law ($\Gamma=2$) absorbed by Galactic column density.  Despite the comparable brightness the flux-resolved spectra differ notably between 1741 and 1742.

The various flux states during each orbit were fitted simultaneously using the best-fit model from Section~\ref{sect:bestfit} as the starting condition.  The warm absorber and distant reflector parameters were fixed to their average values.  Initially, the blurring parameters $q_{in}$ and $R_{b}$ were free, but since neither demonstrated noticeable variability they were fixed to their average values as well.  
The only free parameters were the photon index of the power law continuum, ionisation parameter of the reflector, and the normalisation of both components. 

The spectral variations are well fitted by allowing only these parameters to vary ($\chidof=1.02/3525$ for revolution 
1741 and $\chidof=1.01/2502$ for revolution 1742).
The residuals from the fits are shown in Fig.~\ref{fig:fluxed}.  Some trends resulting from fitting the flux-resolved 
spectra are shown in Fig.~\ref{fig:fluxvar}.  As found when modelling the average spectra, the ionisation parameter 
of the blurred reflector is distinctly higher during revolution 1742.  This is seen in the flux-resolved spectra as 
well; 
however, there does not appear to be a gradual change with increasing flux that might be 
expected (Fig~\ref{fig:fluxvar}a).  We note the telescope does not collect data for $\sim40\ks$ between revolution 
1741 and 1742, which makes it difficult to draw strong conclusions as to when the change in the ionisation parameter occurs.

Other parameters do appear to exhibit predicable behaviour.  The power law slope is seen to steepen with increasing
flux (Fig~\ref{fig:fluxvar}b). The power law flux and reflected flux are also correlated, which would occur if the
primary emitter is brightening intrinsically and illuminating the disc more (Fig~\ref{fig:fluxvar}c).  The
reflection fraction 
($\mathcal{R}$), 
which is taken to be the ratio between the reflected and power law flux between
$20-50\keV$, is also correlated with the reflected flux (Fig~\ref{fig:fluxvar}d).  We  note the relationship between
the two parameters is different at each epoch, which could be due to the simultaneous changes in the power law photon 
indices.

\subsection{Time resolved spectroscopy}
\label{sect:trs}

In order to investigate the model variability over time, spectra were created using $10\ks$ intervals of data resulting in 12 spectra in 1741 and 7 spectra in 1742.  Each spectrum was modelled by allowing only the power law normalisation and slope to vary as well as the reflector normalisation and ionisation.  Fig.~\ref{fig:timeXi} and \ref{fig:timeGam} portray our findings.

\begin{figure*}
\begin{center}
\begin{minipage}{0.48\linewidth}
\scalebox{0.32}{\includegraphics[angle=270]{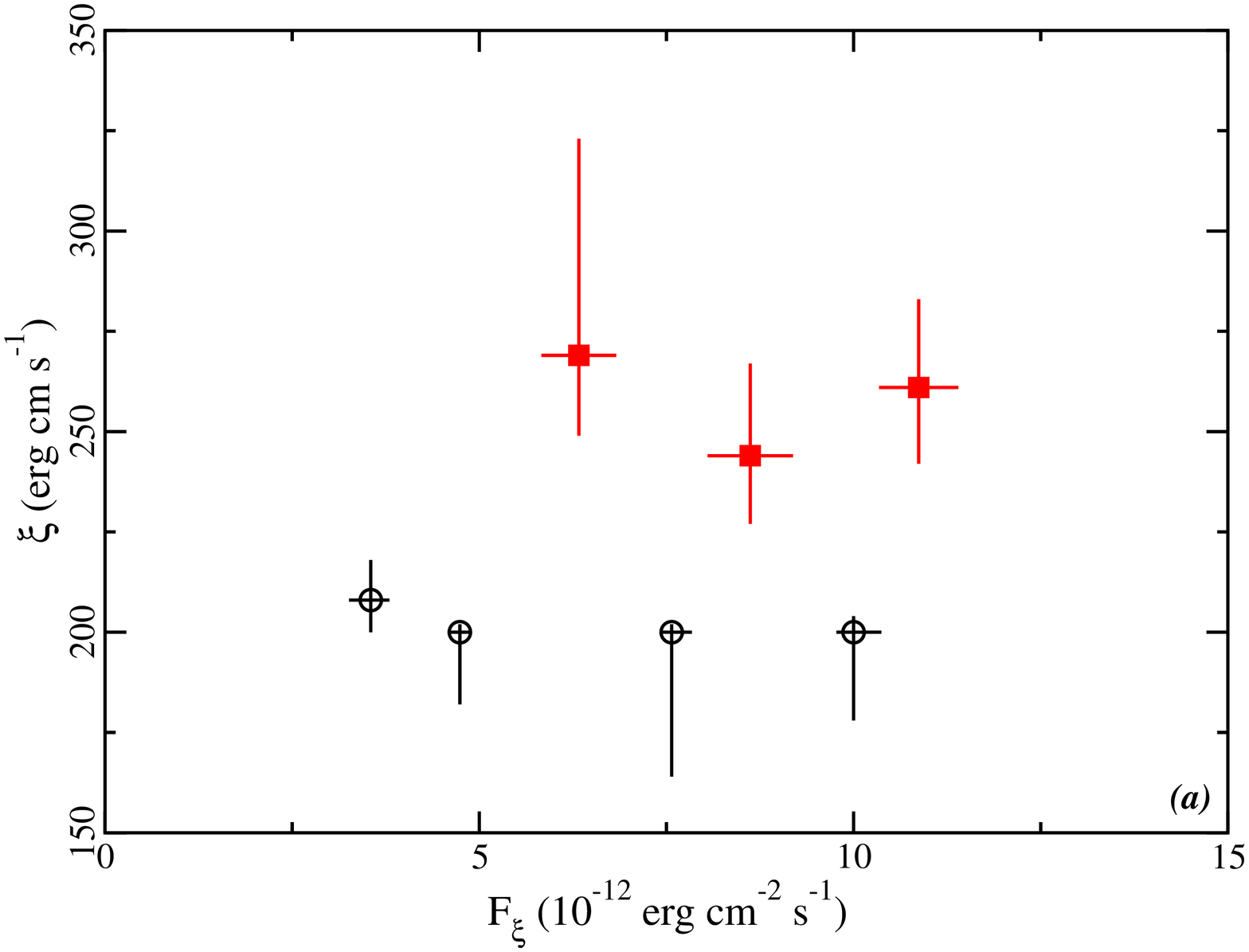}}
\end{minipage}  \hfill
\begin{minipage}{0.48\linewidth}
\scalebox{0.32}{\includegraphics[angle=270]{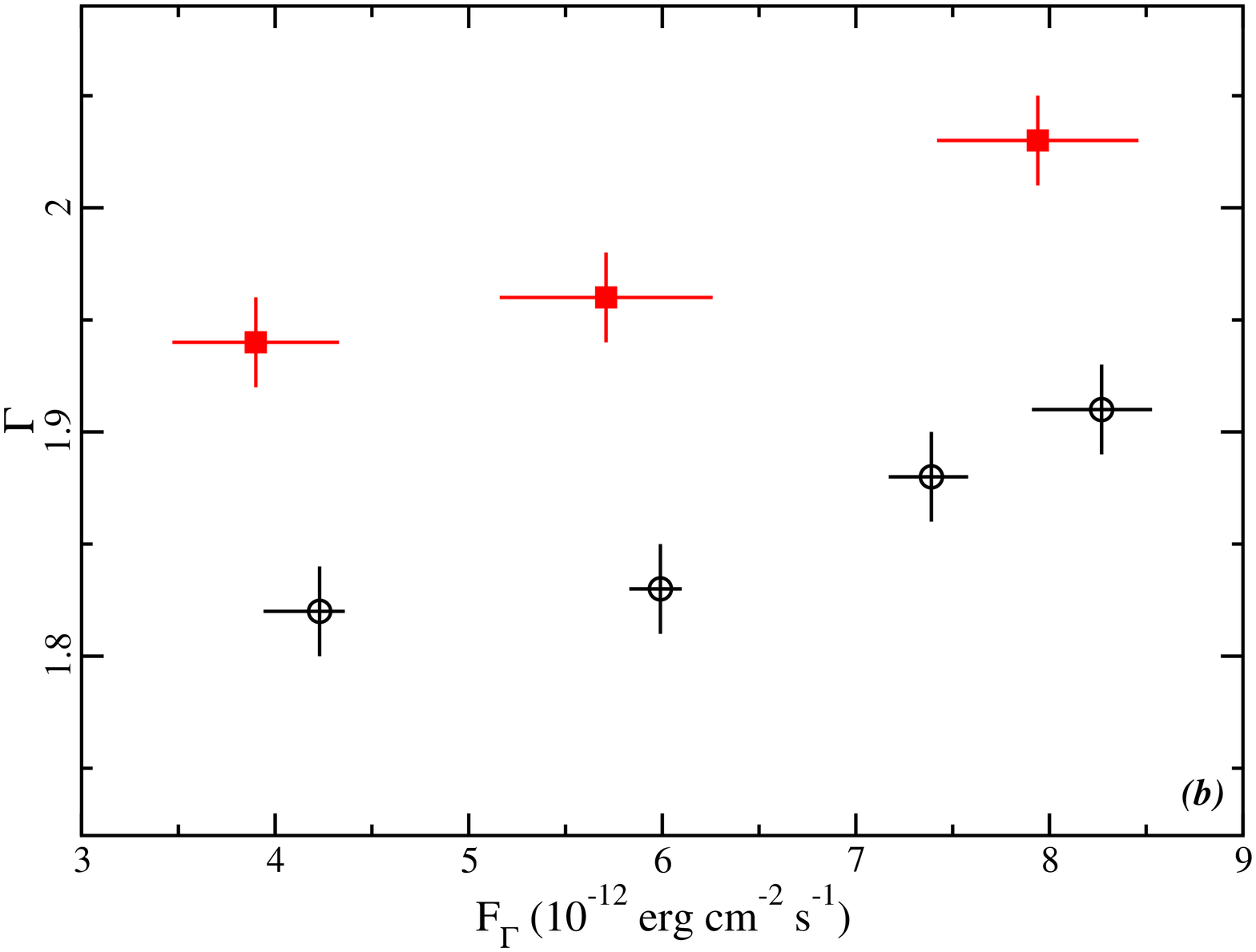}}
\end{minipage}
\begin{minipage}{0.48\linewidth}
\scalebox{0.32}{\includegraphics[angle=270]{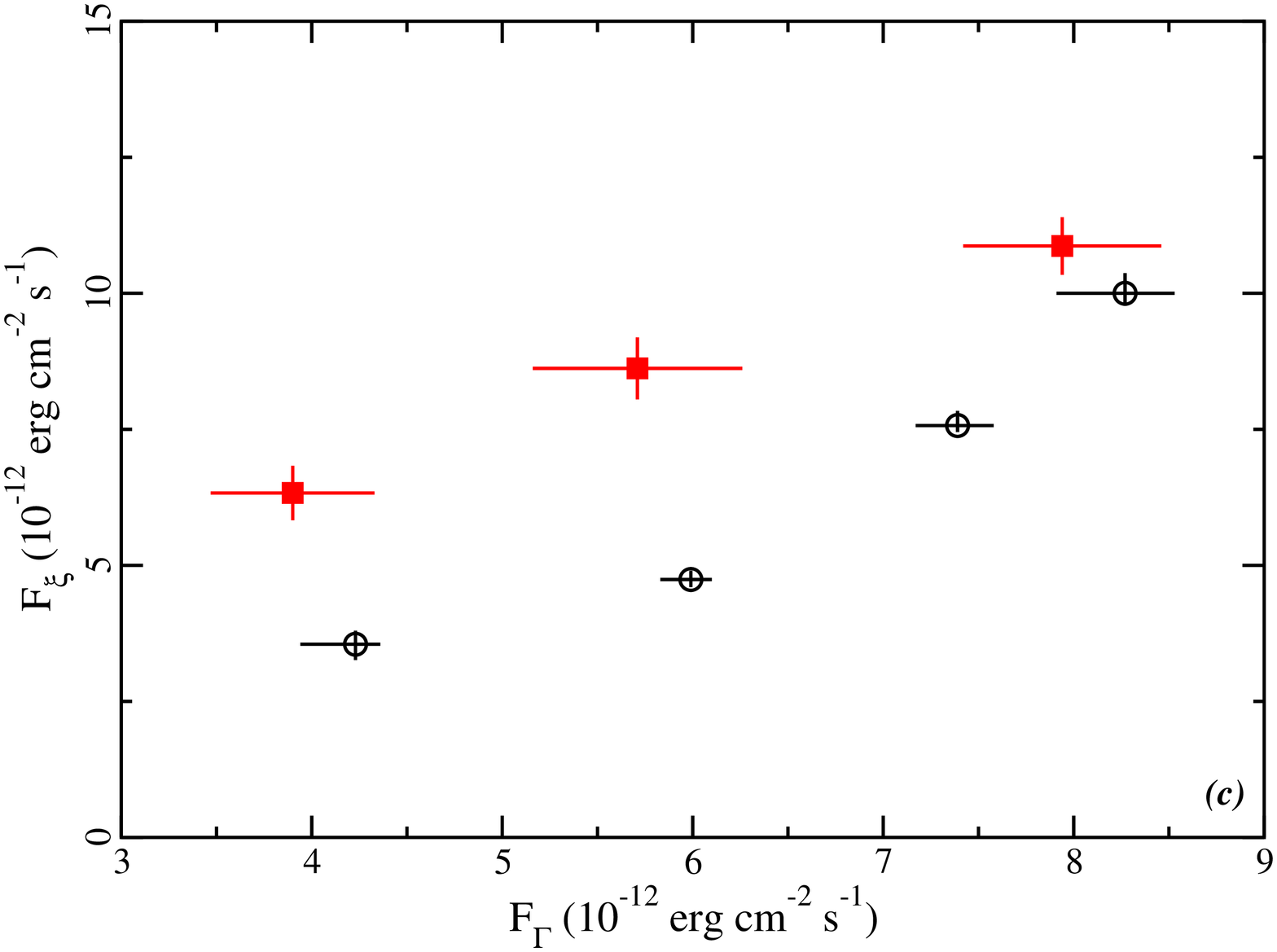}}
\end{minipage}  \hfill
\begin{minipage}{0.48\linewidth}
\scalebox{0.32}{\includegraphics[angle=270]{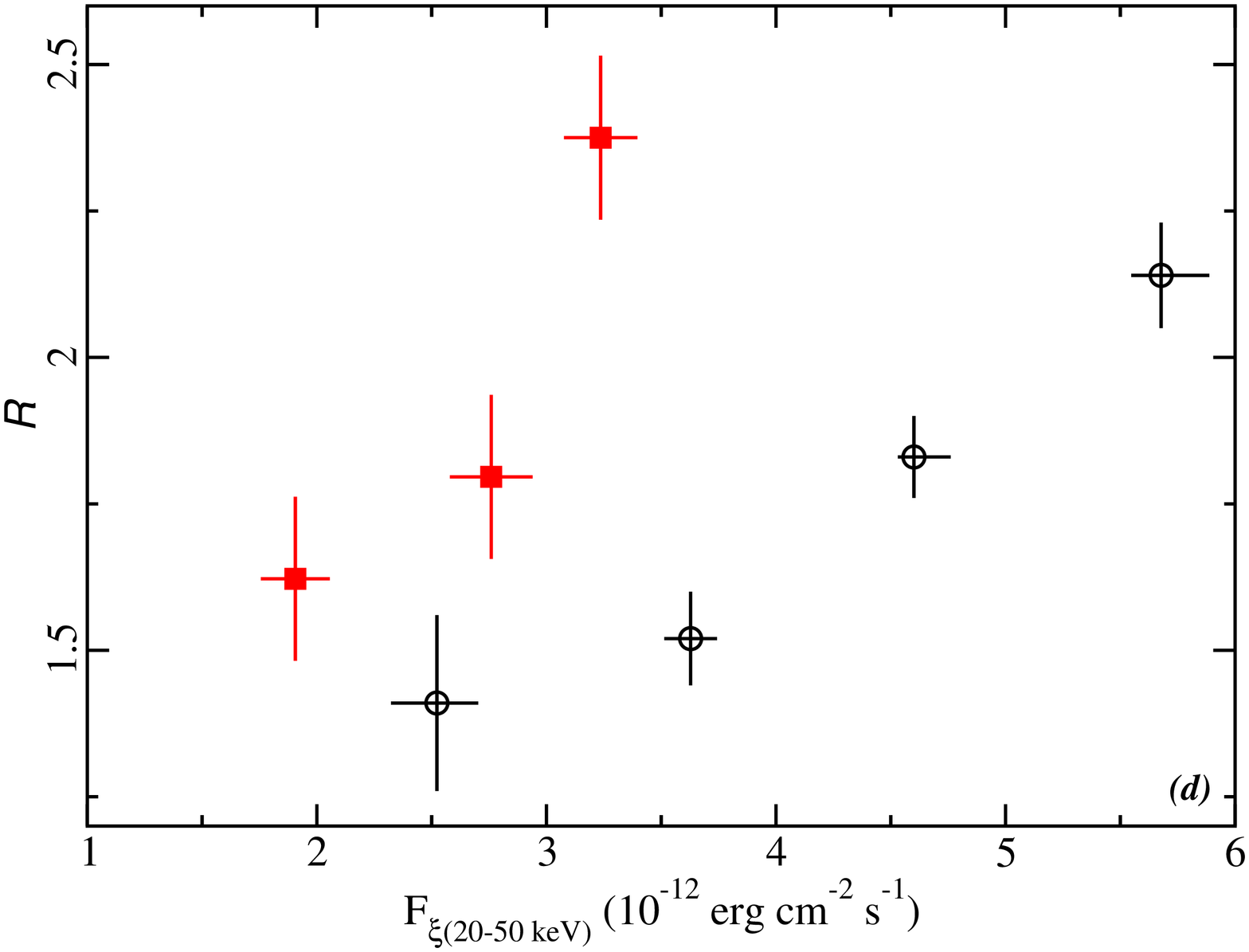}}
\end{minipage}
\end{center}
\caption{The variability of various parameters in different flux states during 1741 (black open circles) and 1742 (red filled squares).  (a) The ionisation parameter of the reflector as a function of the reflected flux ($F_{\xi}$) in the $0.3-10\keV$ band.  (b) The power law photon index as a function of power law flux ($F_{\Gamma}$) in the $0.3-10\keV$ band.  (c) The reflected flux is plotted against power law flux.  (d) The reflection fraction, between $20-50\keV$, is compared to the reflected flux in the same band.}
\label{fig:fluxvar}
\end{figure*}

\begin{figure*}
\begin{center}
\begin{minipage}{0.48\linewidth}
\scalebox{0.32}{\includegraphics[angle=270]{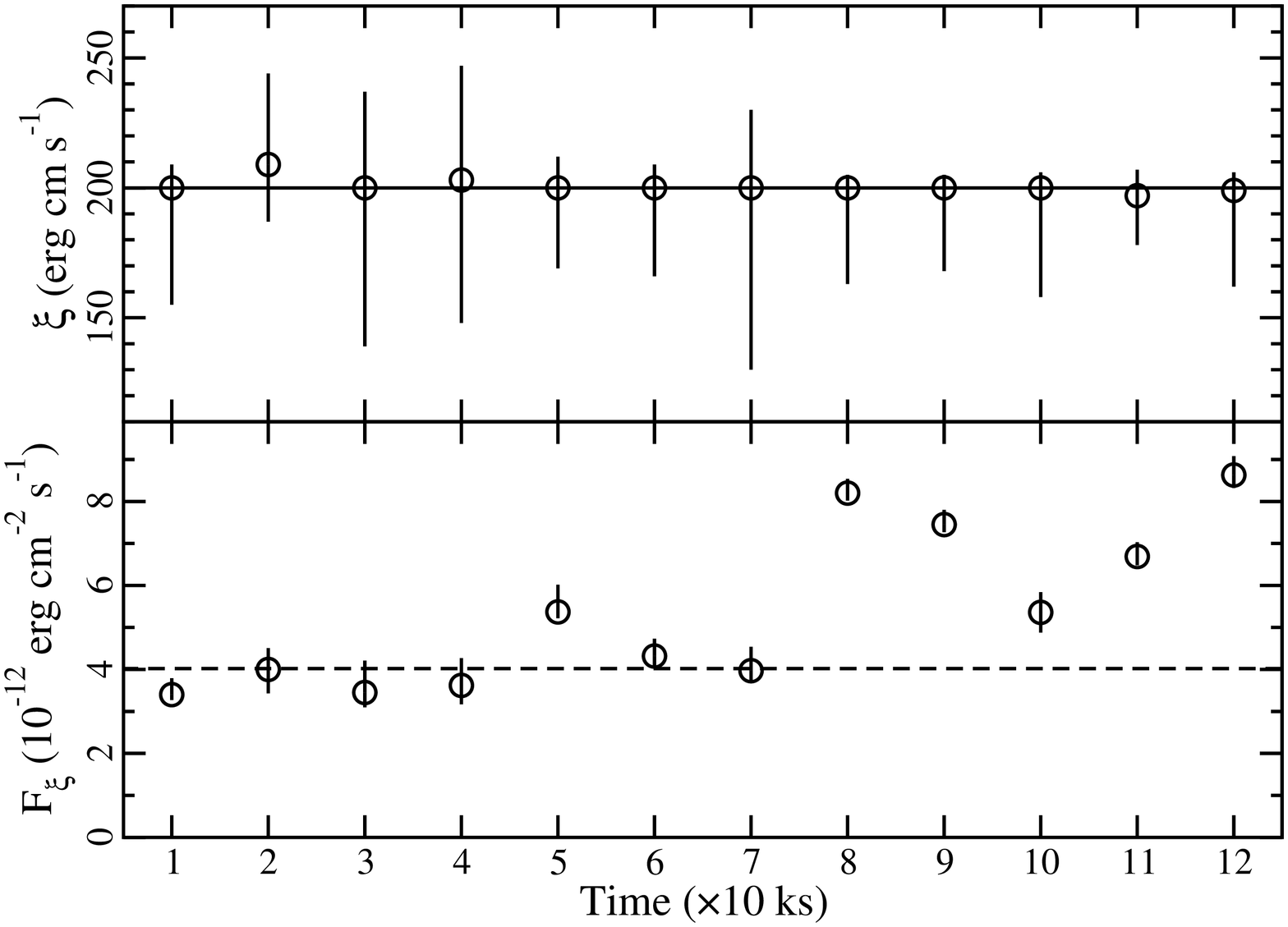}}
\end{minipage}  \hfill
\begin{minipage}{0.48\linewidth}
\scalebox{0.32}{\includegraphics[angle=270]{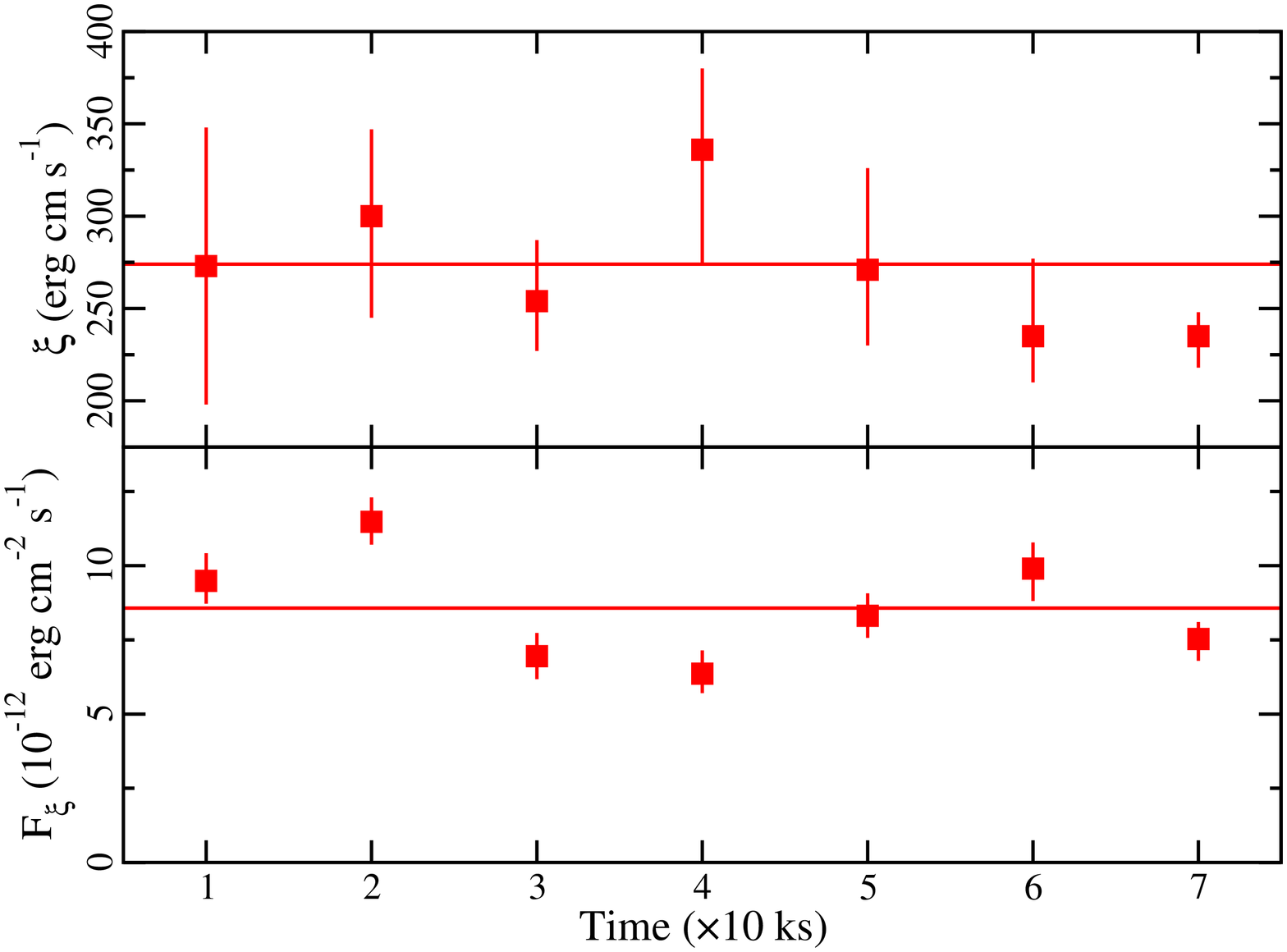}}
\end{minipage}
\end{center}
\caption{The variability of the reflection component is tracked over the entire observation by fitting $10\ks$ resolved spectra during revolution
1741 (left panel) and 1742 (right panel).  The flux (lower panels) and ionisation parameter (upper panels) are compared to the average values (solid lines).  The ionisation parameter appears constant during both observations, but the flux is variable.  The reflected flux during revolution 1741 (lower left panel) is compared to the average during the first $70\ks$ of the observation (dashed line).  The flux appears constant during this segment of the observation.  }
\label{fig:timeXi}
\end{figure*}

\begin{figure*}
\begin{center}
\begin{minipage}{0.48\linewidth}
\scalebox{0.32}{\includegraphics[angle=270]{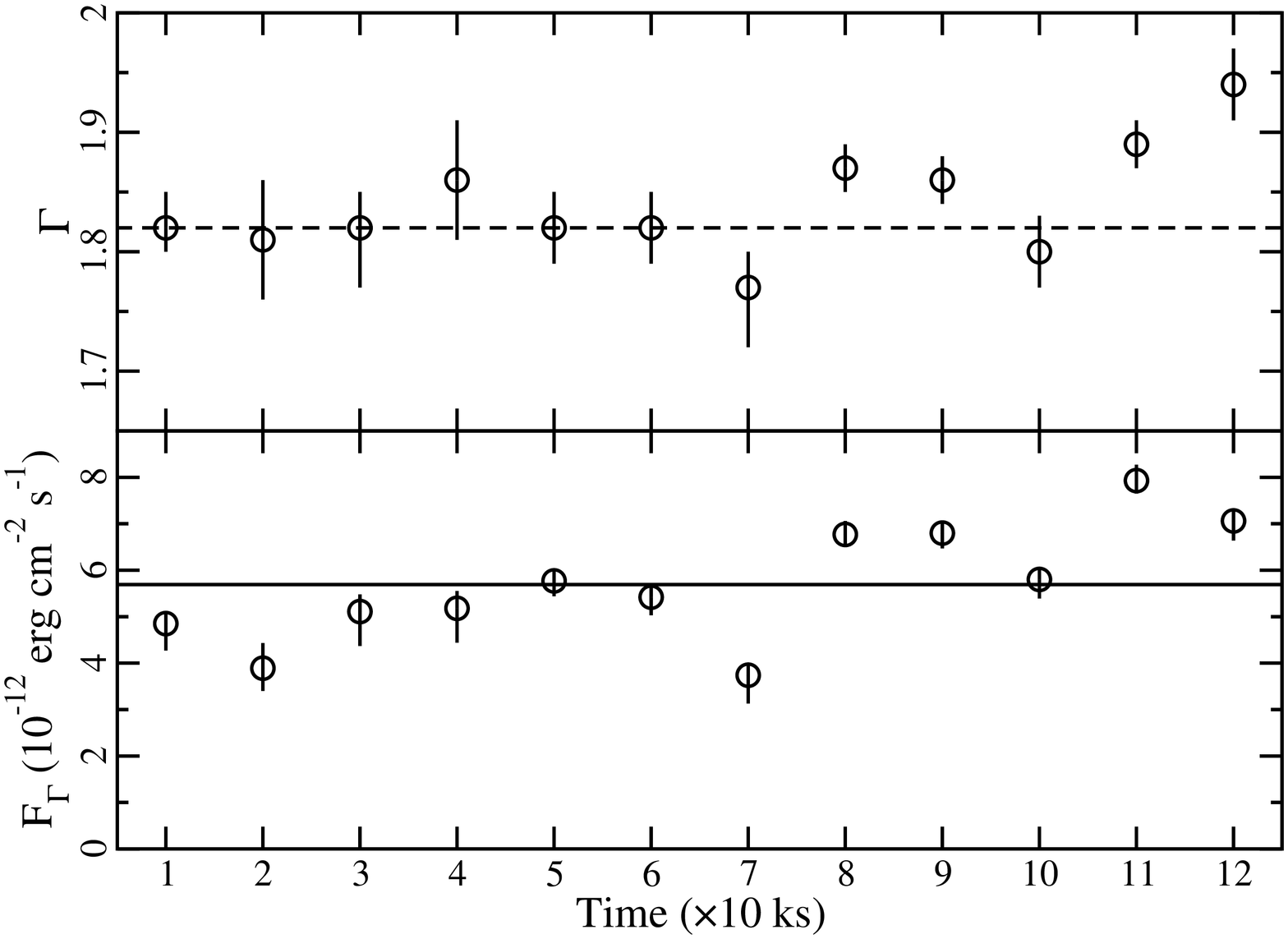}}
\end{minipage}  \hfill
\begin{minipage}{0.48\linewidth}
\scalebox{0.32}{\includegraphics[angle=270]{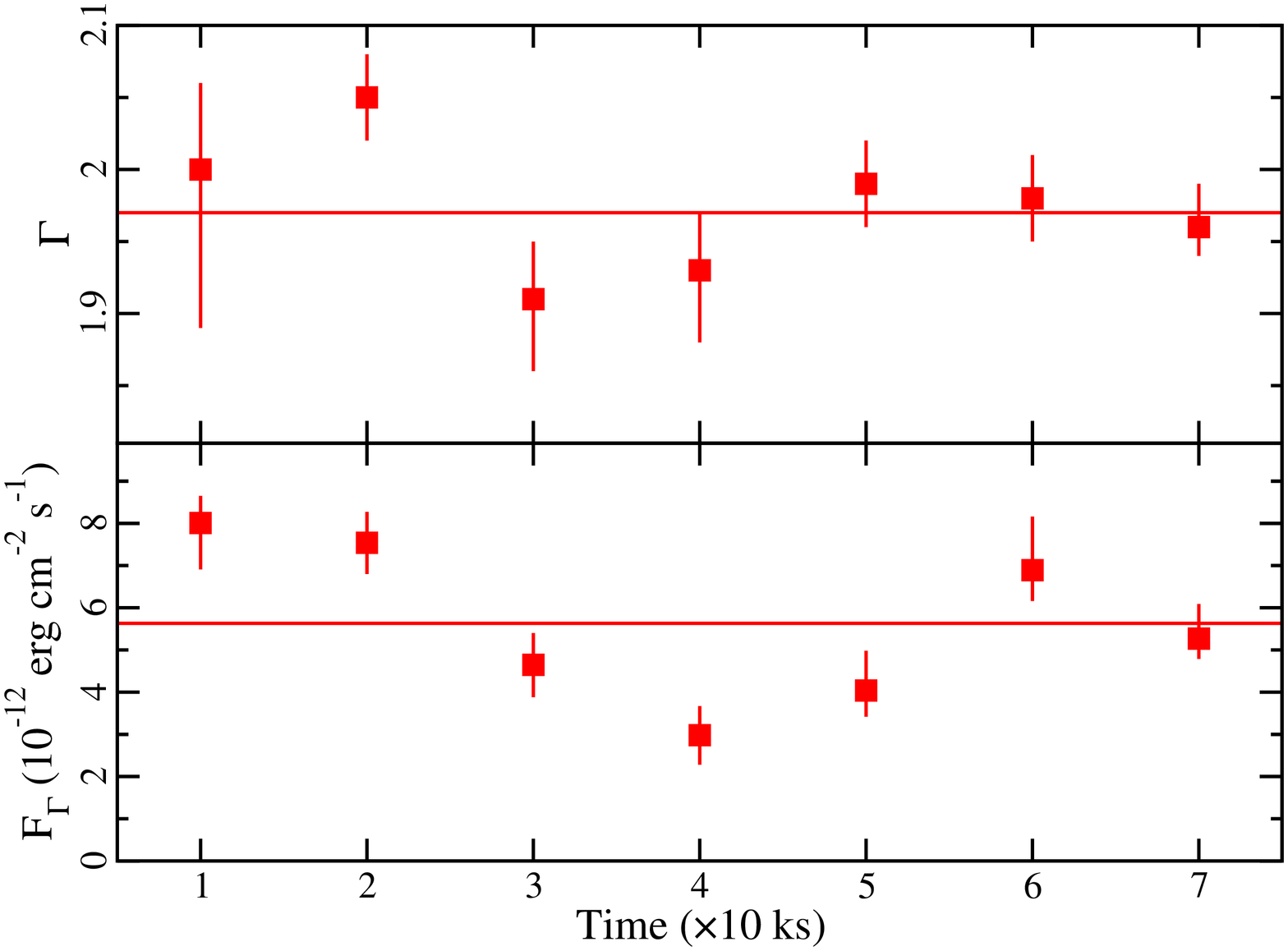}}
\end{minipage}
\end{center}
\caption{The variability of the power law component is tracked over the entire observation by fitting $10\ks$ resolved spectra during revolution
1741 (left panel) and 1742 (right panel).  The flux (lower panels) and photon index (upper panels) are compared to the average values (solid lines) and both parameters appear variable.  The photon index during revolution 1741 (upper left panel) is compared to the average during the first $70\ks$ of the observation (dashed line).  The photon index appears constant during this segment of the observation.  }
\label{fig:timeGam}
\end{figure*}

The average ionisation of the reflector is different at each epochs, but as we see in the top panels of 
Fig.~\ref{fig:timeXi} the ionisation parameter does not vary with the short timescale changes in the 
reflected flux (lower panels of
Fig~\ref{fig:timeXi}).   Similar behavior was seen in the flux-resolved spectra (Fig~\ref{fig:fluxvar}a).

\begin{figure}
\rotatebox{270}
{\scalebox{0.32}{\includegraphics{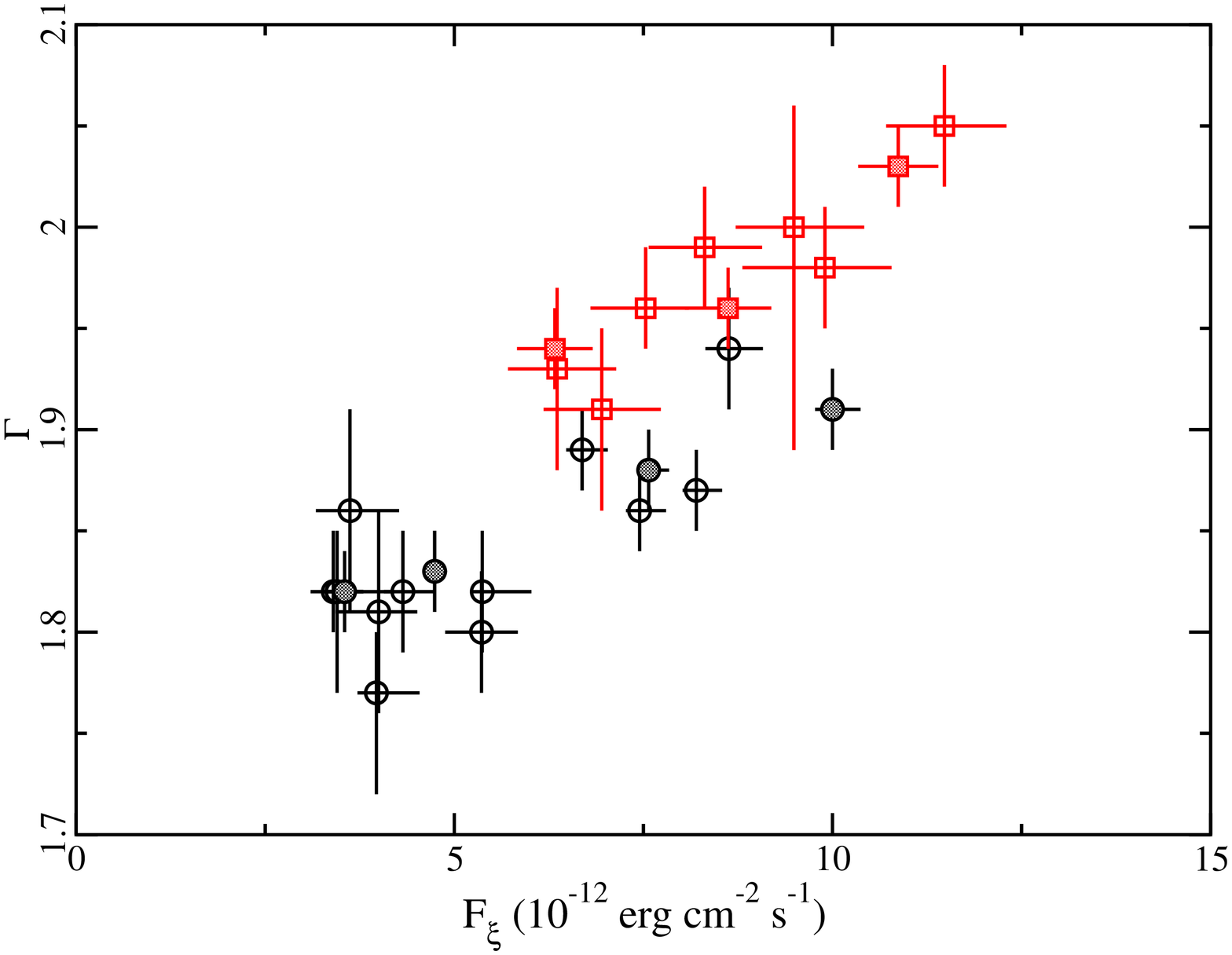}}}
\caption{
The variations of the power law photon index plotted against the variations in the reflected
flux.  The relation between the two parameters is seen in the time-resolved spectra (open symbols) 
as well as in the flux-resolved spectra (shaded symbols).
}
\label{fig:GvsFr}
\end{figure}

The flux of the reflector (lower panels of 
Fig~\ref{fig:timeXi}) is variable over the course of the observations, as is the flux and the photon index of the
power law component (Fig~\ref{fig:timeGam}).  
Of interest, is that during the first $70\ks$ of 1741 the flux of the reflection component is approximately 
constant (Fig.~\ref{fig:timeXi}, lower left panel) as is the shape and flux of the power law component
(Fig~\ref{fig:timeGam} left). 
In general, the reflected flux appears to be well correlated with changes in the power law component.
The specific correlation between the photon index and the reflected flux is shown in 
Fig.~\ref{fig:GvsFr} and exists in both the time-resolved and flux-resolved spectra.  
Since the reflection fraction ($\mathcal R$) increases with reflected flux (Fig~\ref{fig:fluxvar}d), the
observed trend in Fig~\ref{fig:GvsFr} between the shape of the intrinsic power law illuminating the accretion 
disc and the reflected flux is expected
(see e.g. Figure 4 of Ross \& Fabian 2005).

\begin{figure}
\rotatebox{270}
{\scalebox{0.32}{\includegraphics{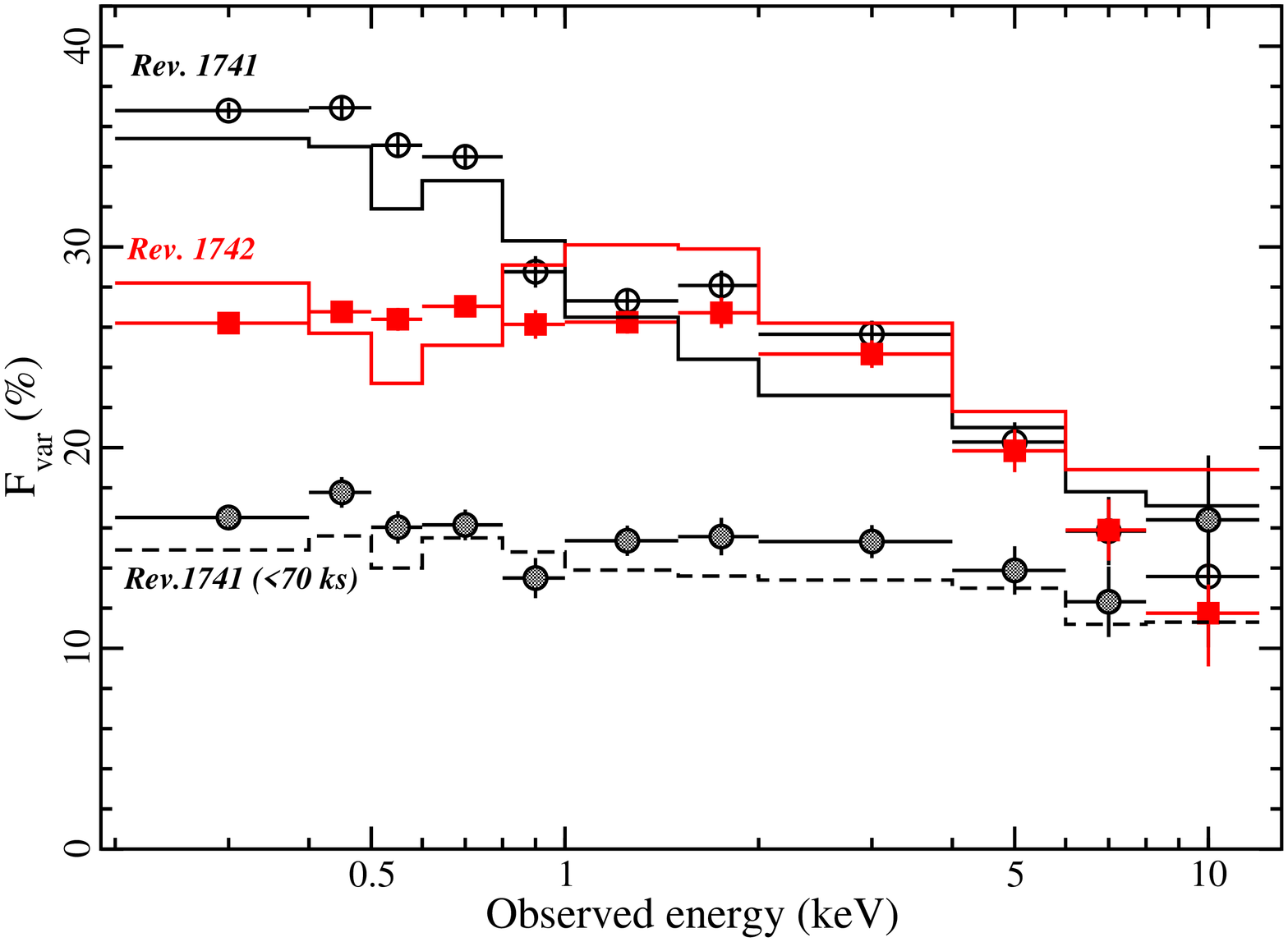}}}
\caption{
The fractional variability as of function of energy is calculated for observation 1741 (black open circles) and 1742 (red filled squares), as well as for the first $70\ks$ of observations 1741 (black, shaded circles).  The models described in Section~\ref{sect:trs} are overplotted and show reasonable agreement with the data.
}
\label{fig:fvar}
\end{figure}

The variability behaviour also describes the appearance of the fractional variability ($\fvar$) spectrum of \mrk335\ at each epoch (Fig.~\ref{fig:fvar}).  The $\fvar$ in each energy band is calculated with light curves in $500\s$ bins following Edelson \et (2002) and uncertainties are estimated following Ponti \et (2004).  
There is clearly a difference below $\sim 1\keV$ in the $\fvar$ spectra between the two epochs with 1741 showing larger variations at lower energies.  The $\fvar$ spectrum is also calculated for only the first $70\ks$ of revolution 1741 (shaded, black circles), showing much smaller variations over this reduced period.    Each $\fvar$ spectrum can be reasonably depicted by the variations found in Fig.~\ref{fig:timeXi} and ~\ref{fig:timeGam}.  The first $70\ks$ of 1741 are described by variations in only the power law normalisation, while to describe the average $\fvar$ in 1741 (black open circles) variability in the remaining $\sim 50\ks$ requires additional changes in the power law slope and reflector normalisation.  The variability in 1742 also requires changes in the power law shape and normalisation, and in the reflector normalisation.

\section{Discussion} 

Blurred reflection is expected when the inner accretion disc extends down toward the radius of marginal stability around a black
hole.  Many of the concerns raised over blurred reflection are now coming into focus with recent X-ray data.  
The discovery of broadened
\fela\ and short lags (e.g. Fabian \et 2009; Zoghbi \et 2010) suggest the soft excess in AGN is due to reflection and that
it arises from reprocessing of the primary continuum.  The broad \feka\ line is also seen in black hole binaries (BHB) 
(e.g. Miller \et 2009) showing that reflection is ubiquitous in black hole systems.  The similarities in the \feka\ profile seen 
in AGN and BHB (Walton \et 2012b) argues strongly in favour of the reflection interpretation.  Moreover, the evidence for 
blurred reflection is not limited to unique systems or specific lines-of-sight.  Evidence of reflection is seen in most systems
where there is a direct view of the central regions (e.g. Crummy \et 2005; Zoghbi \et 2012; Walton \et 2012a; De Marco \et 2011).

The \xmm\ observations of \mrk335\ analyzed here provide a view of the AGN that is distinct from previous observations.  Firstly, by the presence of a complex warm absorber (analyzed by L12), and secondly through the significant and rapid spectral variability.   Over the course of the $200\ks$ observation the source becomes notably softer as it brightens.  When examining the light curve from the  long-term \swift\ monitoring campaign (G12), we find that these \xmm\ observations catch \mrk335\ in rapid transition from a low-flux to a high-flux state (G12). The X-ray continuum is well described with a primary power law emitter and blurred reflection from inner region of the accretion disc around a rapidly spinning black hole.  The blurred reflection model also accurately describes the variability assuming that the continuum components are allowed to vary.  Specifically, the photon index and power law normalisation as well as the reflector ionisation and normalisation are changing.

Though the behaviour is complex, the variability in the parameters is largely consistent with what is expected.  The overall changes appear to be driven by the pivoting of the power law component (i.e. the primary emitter).  
The power law flux increases as the spectrum steepen, in line with expectations (e.g. Haardt \et 1997). 
Assuming the distance between the primary emitter and the disc does not change, the flux and ionisation of the disc will increase accordingly, as is seen.

The changes in power law photon index occur very rapidly.  During the first $70\ks$ of the observation  \mrk335\ did not exhibit any notable spectral variability.  The apparent change in the photon index, specifically steepening from $\Gamma\approx1.83$ to $\Gamma\approx1.98$, occurs over $\sim50\ks$ in the rest frame.  A scenario in which the primary emitter could be moving at mildly relativistic velocities with respect to the disc (e.g. Reynolds \& Fabian 1997; Beloborodov 1999) could be invoked to drive the changes in $\Gamma$.  This interpretation has been advocated for other AGN and NLS1s (e.g. Gallo \et 2007, 2011; Miniutti \et 2010).  However, while light bending is certainly prominent, the rather constant emissivity profile and break radius obtained in our spectral fitting of \mrk335\ suggest that the blurring parameters are not changing as would be expected if the primary source is varying its distance from the disc.  

Fabian \et (2012) suggest a double power law continuum to describe the low state X-ray spectrum of Cyg X-1.  There they suggest the steeper component originates at the base of the jet and is responsible for the reflection with the harder component located farther up the jet.  This scenario could perhaps be expanded to describe the observations in \mrk335.  One could imagine multiple jet components coming together at the base of a jet located close to the disc, for example a jet ``blob'' that does not obtain the escape velocity and falls back downstream toward the base (e.g.  Ghisellini \et 2004).  The coming together of two components of differing densities could produce the changes seen in the power law and ultimately in the reflection.  Interestingly, a double power law continuum was suggested for the blurred reflection interpretation of the very low flux state observation of \mrk335\ (Grupe \et 2008).
\mrk335\ is certainly radio-quiet, but not radio silent ($R =L_{\nu}(5GHz)/L_{\nu}(B)  \approx 0.25$) so the possibility of some jet emission does exist.

\section{Conclusions } 

The most recent \xmm\  observations of \mrk335\ catch the AGN in a complex, intermediate flux state.  Both the X-ray continuum and timing properties of the NLS1 can be described in a self-consistent manner adopting a blurred reflection model with no need to invoke partial covering.  The rapid spectral variability appear to be driven by changes in the spectral shape of the primary emitter that is illuminating the inner accretion disc.   


\section*{Acknowledgments}

The \xmm\ project is an ESA Science Mission with instruments
and contributions directly funded by ESA Member States and the
USA (NASA).  We are grateful to the \xmm\ observing team for preparing and
activating the ToO.  
We thank the referee for comments that improved the clarity of the paper.
DG acknowledges support from NASA contracts NNX07AH67G and NNX09AN12G




\bsp
\label{lastpage}
\end{document}